\documentclass[letterpaper,twoside,10pt]{article}
\usepackage{amssymb}
\usepackage{latexsym}
\usepackage{verbatim}
\usepackage{epsfig}
\usepackage{fancyheadings}
\usepackage{pstricks,pst-node}
\usepackage{rotating,graphicx}
\newlength{\figurewidth}
\newcommand{\capitem}[1]{\caption{\textsf{#1}}}

\newcommand{\GMab}{(C \Gamma^\mu)_{\alpha\beta}}

\newcommand{\be}{\begin{equation}}
\newcommand{\ee}{\end{equation}}
\newcommand{\ble}[1]{\begin{equation} \label{#1}}
\newcommand{\bae}{\begin{eqnarray}}
\newcommand{\eae}{\end{eqnarray}}
\def\nn{\nonumber}
\newcommand{\ff}{\nn \\}
\def\fe{& = &}

\def\eg{\hbox{\it e.g.}}
\def\etc{\hbox{\it etc.}}
\def\ie{\hbox{\it i.e.}}
\newcommand{\calF}{\mathcal{F}}
\newcommand{\calG}{\mathcal{G}}

\newcommand{\calL}{\mathcal{L}}

\newcommand{\calO}{\mathcal{O}}

\begin{document}
%%%%%%%%%%%%%%%%%%%%%%%%%%%%%%%%%%%%%%%%%%%%%%%%%%%%%%%%%%%%%%
%%%%%%%%%%%%%%%%%%%%%%%%%%%%%%%%%%%%%%%%%%%%%%%%%%%%%%%%%%%%%%
%%%%%%%%%%%%%%%%%%%%%%%%%%%%%%%%%%%%%%%%%%%%%%%%%%%%%%%%%%%%%%
% Titlepage
%%%%%%%%%%%%%%%%%%%%%%%%%%%%%%%%%%%%%%%%%%%%%%%%%%%%%%%%%%%%%%
\begin{titlepage}
\vspace*{-1cm}
\begin{flushright}
\textsf{ICN-UNAM-01/01}
\\
\mbox{}
\\
\textsf{Feb 15, 2001}
\\[3cm]
\end{flushright}
%%%%%%%%%%%%%%%%%%%%%%%%%%%%%%%%%%%%%%%%%%%%%%%%%%%%%%%%%%%%%%
%%%%%%%%%%%%%%%%%%%%%%%%%%%%%%%%%%%%%%%%%%%%%%%%%%%%%%%%%%%%%%
%%% TITLE, AUTHORS
%%%%%%%%%%%%%%%%%%%%%%%%%%%%%%%%%%%%%%%%%%%%%%%%%%%%%%%%%%%%%%
%%%%%%%%%%%%%%%%%%%%%%%%%%%%%%%%%%%%%%%%%%%%%%%%%%%%%%%%%%%%%%
%\begin{center}
%%%%%%%%%%%%%%%%%%%%%%%%%%%%%%%%%%%%%%%%%%%%%%%%%%%%%%%%%%%%%%
\renewcommand{\thefootnote}{\fnsymbol{footnote}}
\begin{LARGE}
\bfseries{\sffamily Stability of Lie Superalgebras and Branes}
\end{LARGE}

\noindent \rule{\textwidth}{.6mm}

\vspace*{1.6cm}

\noindent
\begin{large}%
\textsf{\bfseries%
Chryssomalis Chryssomalakos
}
\end{large}

%\vspace*{.1cm}

\phantom{XX}
\begin{minipage}{.8\textwidth}
\begin{it}
\noindent Instituto de Ciencias Nucleares \\
Universidad Nacional Aut\'onoma de M\'exico\\
Apdo. Postal 70-543, 04510 M\'exico, D.F., MEXICO \\
\end{it}
\texttt{chryss@nuclecu.unam.mx
\phantom{X}}
\end{minipage}
\\

\vspace*{3cm}
%%%%%%%%%%%%%%%%%%%%%%%%%%%%%%%%%%%%%%%%%%%%%%%%%%%%%%%%%%%%%%
%%% ABSTRACT
%%%%%%%%%%%%%%%%%%%%%%%%%%%%%%%%%%%%%%%%%%%%%%%%%%%%%%%%%%%%%%
\noindent
\textsc{\large Abstract: }
The algebra of the generators of translations in superspace is
unstable, in the sense that infinitesimal perturbations of its
structure constants lead to non-isomorphic algebras. We show how
superspace extensions remedy this situation (after arguing that
remedy is indeed needed) 
and review the benefits reaped in the description of 
branes of all kinds in the presence of the extra dimensions.   
%%%%%%%%%%%%%%%%%%%%%%%%%%%%%%%%%%%%%%%%%%%%%%%%%%%%%%%%%%%%%%
\end{titlepage}
\setcounter{footnote}{0}
\renewcommand{\thefootnote}{\arabic{footnote}}
\setcounter{page}{1}
%%%%%%%%%%%%%%%%%%%%%%%%%%%%%%%%%%%%%%%%%%%%%%%%%%%%%%%%%%%%%%%%%
%%%%%%%%%%%%%%%%%%%%%%%%%%%%%%%%%%%%%%%%%%%%%%%%%%%%%%%%%%%%%%%%%
%\noindent \rule{\textwidth}{.5mm}
%
%\tableofcontents
%
%\noindent \rule{\textwidth}{.5mm}
%%%%%%%%%%%%%%%%%%%%%%%%%%%%%%%%%%%%%%%%%%%%%%%%%%%%%%%%%%%%%%%%
%%%%%%%%%%%%%%%%%%%%%%%%%%%%%%%%%%%%%%%%%%%%%%%%%%%%%%%%%%%%%%%%
%%%%%%%%%
%%%%%%%%% TEXT BEGINS
%%%%%%%%%
%%%%%%%%%%%%%%%%%%%%%%%%%%%%%%%%%%%%%%%%%%%%%%%%%%%%%%%%%%%%%%%%
%%%%%%%%%%%%%%%%%%%%%%%%%%%%%%%%%%%%%%%%%%%%%%%%%%%%%%%%%%%%%%%%%
\section{Introduction}
\label{Intro}
%%%%%%%%%%%%%%%%%%%%%%%%%%%%%%%%%%%%%%%%%%%%%%%%%%%%%%%%%%%%%%%%
%%%%%%%%%%%%%%%%%%%%%%%%%%%%%%%%%%%%%%%%%%%%%%%%%%%%%%%%%%%%%%%%%
The aim of physical theories is the simplest possible 
description of Nature within the uncertainties afforded by
experiments. The symmetry aspects of this description are 
usually codified in an underlying Lie algebra, around which
the theory is built. The structure
constants of the algebra parameterize the theory and are, in
principle, measurable quantities --- they often emerge as the
fundamental constants of the theory. Given that our knowledge of
these parameters can only be approximate, physical theories that
do not change qualitatively under small perturbations of their
parameters have more chances of wide applicability. This observation 
may not be worthy of an axiom status 
but it does provide a sensible criterion
about which regions of parameter space might be most promising
for theory hunting. 

Lie algebras come in two varieties: {\em stable}%
\footnote{%
the term {\em rigid} is often used by
mathematicians as a synonym.%
} 
and {\em unstable}. The former are
isomorphic to all Lie algebras in their vicinity (in a sense to
be made precise later on) while the latter are infinitesimally
close to qualitatively different (\ie \ non-isomorphic) algebras. 
To specify the content of a physical theory relying on an unstable
Lie algebra, an infinitely fine tuning of the values of its
parameters is necessary. 
It makes therefore sense, when confronted with such a
theory, to look for generalizations that are immune to
perturbations, \ie, are endowed with a
definite structure despite the fuzziness in their parameters. 
This point of view has an already long history, and not only in
physics. It permeates the work of Thom~\cite{Tho:75} on
morphogenetic mechanisms in the
seventies, which in turn draws on Smale's~\cite{Sma:67} strong 
advocation, as early 
as in the sixties, of a similar principle in the study of 
non-linear dynamics. In physics, it has been pointed out by
several authors~\cite{Fla:82,Fad:88,Bay.Fla.Fro.Lic.Ste:78,Vil:94} 
that the
kinematics of special relativity can be obtained as a
stabilization of the Galilean algebra, an example that we
also present in the sequel, as well as that the quantization
of classical mechanics can be described as the
transition from an (infinite dimensional) unstable Lie algebra
of functions on phase space to a stable one. In either case, the
mathematical process involved has been formalized in the theory of
deformations of Lie algebras, developed in the sixties by
Gersternhaber~\cite{Ger:64}, Nijenhuis and
Richardson~\cite{Nij.Ric:67} and others. 

On a different front, alluded to by the second half of the title,
recent years have witnessed an explosion of interest in extended
objects, collectively referred to as branes, for which a variety
of action functionals has by now been proposed and studied. 
The matching of the number of bosonic and fermionic degrees 
of freedom in these actions is guaranteed by the so-called
$\kappa$-symmetry, which in turn necessitates the introduction 
of a WZ term. Already in the case of the superstring
action of Green and Schwarz, a problem arises by the fact that
the WZ term is not manifestly supersymmetric, but rather,
transforms
by a total derivative under general translations in superspace.
Siegel~\cite{Sie:94} showed that by augmenting superspace by a 
new fermionic variable, one could write a truly 
invariant WZ term for the action, consisting of the pull-back on
the worldvolume of
an invariant 2-form defined on the {\em extended} superspace. At 
the algebra level this amounts to a central 
extension by a fermionic generator which transforms as a spinor
under the Lorentz group. It was thus shown that, once
extended objects are considered, further extensions of standard
superspace might provide a natural background for the description
of their dynamics. Following Siegel's work, Bergshoeff and
Sezgin~\cite{Ber.Sez:95}
proposed extended supersymmetry algebras that lend themselves to
the description of $p$-branes. The zoo of extended objects (and
of corresponding algebras) has been
enriched since then with the appearance of
$D$-branes~\cite{Pol:95} (and,
later, $L$-branes~\cite{How.Rae.Rud.Sez:98}),
the actions of which differ qualitatively from the standard brane
action in that they involve fields defined directly on the
worldvolume. 

The plethora of extended objects
mentioned above, combined with the 
disparity in the characteristics of their actions, motivated the
systematic study of superalgebra extensions undertaken 
in~\cite{Chr.Azc.Izq.Bue:00}. It was shown there how genuinely
invariant actions can be found on suitably extended superspaces,
recovering the results of~\cite{Ber.Sez:95}. Even more
tantalizingly, the 
Born-Infeld worldvolume fields were shown to come from pull-backs
from these extended superspaces.
Our aim here is to show how
these results reinforce the particular point of view
advocated above and to argue that they should be added to the list of
developments illuminated by stability considerations.
Sec.{}~\ref{Defo} starts with a cursory look at the formal
deformation theory of Lie algebras,
including the
example of the Galilean algebra. 
Then branes enter, in Sec.{}~\ref{QiA}, and
some of the problems arising in the standard superspace formulation
are pointed out. Sec.~\ref{Ext} shows how to extend superspace so
that the resulting algebra of the generators of supertranslations
is stabilized. Sec.~\ref{Appli} looks at the applications to
standard $p$-branes and $D$-branes mentioned above.
We contemplate on
future possible directions, adhering to the stability theme, in
the epilogue.   
%%%%%%%%%%%%%%%%%%%%%%%%%%%%%%%%%%%%%%%%%%%%%%%%%%%%%%%%%%%%%%%%%
%%%%%%%%%%%%%%%%%%%%%%%%%%%%%%%%%%%%%%%%%%%%%%%%%%%%%%%%%%%%%%%%%
\section{Deformations} 
\label{Defo}
%%%%%%%%%%%%%%%%%%%%%%%%%%%%%%%%%%%%%%%%%%%%%%%%%%%%%%%%%%%%%%%%%
%%%%%%%%%%%%%%%%%%%%%%%%%%%%%%%%%%%%%%%%%%%%%%%%%%%%%%%%%%%%%%%%%
Given an $n$-dimensional real\footnote{%
{\em Lie algebra} will always mean {\em real Lie algebra} in what
follows.}
Lie algebra $\calG$, with generators $T_A$,
$A=1, \ldots ,n$. $\calG$ is specified completely by its (real) 
structure constants ${f_{AB}}^C$. The latter are subject to two
conditions: antisymmetry in the lower two indices and the Jacobi
identity 
\bae
\label{antiJI}
{f_{AB}}^C \fe - {f_{BA}}^C
\, ,
\ff
  {f_{AR}}^S {f_{BC}}^R 
+ {f_{BR}}^S {f_{CA}}^R 
+ {f_{CR}}^S {f_{AB}}^R 
\fe 0
\, .
\eae
Relaxing for the moment the latter, one is
left with $N(n)=n^2(n-1)/2$ arbitrary constants. Consider now
the space $\mathbb{R}^N$, with each of the $f$'s ranging along an axis. 
For each value of $(A,B,C,S)$,~(\ref{antiJI}) describes a
quadratic hypersurface in this space. The intersection of 
these hypersurfaces is
the space $\calL_n$ of all possible $n$-dimensional Lie algebras
--- we sketch it as a surface in Fig.~1. 
%%%%%%%%%%%%%%%%%% FIGURE
\setlength{\figurewidth}{.9\textwidth}
\begin{figure}
\begin{pspicture}(0mm,0mm)(\figurewidth,.63\figurewidth)
\setlength{\unitlength}{.25\figurewidth}
\psset{xunit=.25\figurewidth,yunit=.25\figurewidth,arrowsize=1.5pt 3}
\centerline{\raisebox{1.05\totalheight}{%
\includegraphics[angle=270,width=.83\figurewidth]{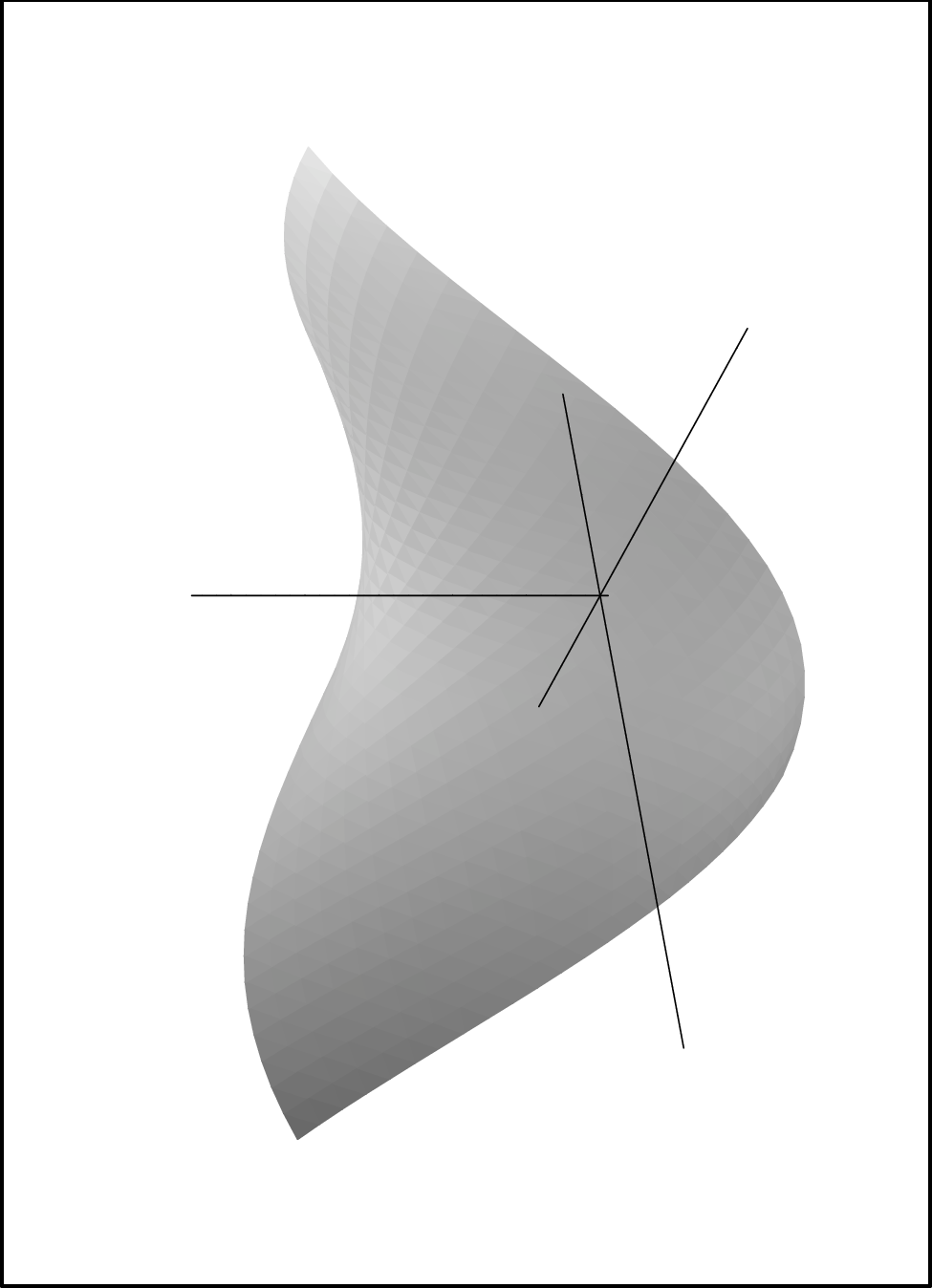}}}
\pscustom[fillstyle=solid,fillcolor=gray]{%
\psccurve(-2.6,1.67)(-2.4,1.54)(-2.12,1.4)(-2.2,1.3)(-2.46,1.33)}
\psdots[dotscale=1.3](-2.5,1.5)(-2.35,1.4)(-1.65,1.35)
\pscurve[linewidth=.3mm]{-}(-1.75,1)(-1.65,1.35)(-1.35,1.48)
\psline[linewidth=.3mm]{->}(-2.5,1.5)(-2.35,1.4)
\psline[linewidth=.3mm]{->}(-1.65,1.35)(-1.83,1.41)
\put(-2.9,1.7){\makebox[0cm][r]{{\Large $\calL_n$}}}
\put(-2.5,1.53){\makebox[0cm][r]{$P$}}
\put(-2.33,1.35){\makebox[0cm][l]{$P_M$}}
\put(-1.63,1.24){\makebox[0cm][l]{$Q$}}
\put(-1.80,1.43){\makebox[0cm][r]{$\psi^{(2)}_1$}}
\rput(-2.6,1.18){\rnode{A}{$GL(n)$ orbit of $P$}}
\rput(-2.45,1.4){\rnode{B}{}}
\nccurve[angleA=45,angleB=-110,linewidth=.18mm]{->}{A}{B}
\rput(-1.55,1.85){\rnode{A}{$GL(n)$ orbit of $Q$}}
\rput(-1.36,1.49){\rnode{B}{}}
\nccurve[angleA=-45,angleB=90,linewidth=.18mm]{->}{A}{B}
\end{pspicture}
%%%%%%%%%%%%%%%%%%%%%%% Caption
\capitem{\textsf{The space $\calL_n$ of $n$-dimensional Lie
algebras. $P$ is surrounded by equivalent points and hence,
$\calG_P \sim \calG_{P_M}$, for all $P_M$ sufficiently close to
$P$. In contrast, in the
tangent space of $\calL_n$ at $Q$, there are directions that lead
outside of the orbit $\calO(Q)$. $Q$ will move along these 
directions when
$\psi^{(2)}_1$ in~(\ref{defcom}) is a non-trivial element of
$H^2(\calG_Q)$.}}
\label{allLA}
\end{figure}
%%%%%%%%%%%%%%%%%% FIGURE
Referring to
this figure, consider the point $P$ on $\calL_n$ --- it corresponds
to the Lie algebra $\calG_P$, whose structure constants are given by
the coordinates of $P$. Under a linear redefinition of the 
generators via a $GL(n)$ matrix $M$,
\ble{linredef}
T'_A = {M_A}^B T_B
\, ,
\ee
the structure constants transform as
\ble{sctran}
{f'_{AB}}^C= {M_A}^R {M_B}^S {(M^{-1})_U}^C {f_{RS}}^U
\, ,
\ee
and $P$ moves to $P_M$. Clearly, no new physics is to be expected
from such a redefinition, $\calG_P$ and $\calG_{P_M}$ being
isomorphic. What we are really interested in then,
from a physical point of view, is
not $\calL_n$ itself, but, rather, the equivalence classes into which
$\calL_n$ splits under the above action of $GL(n)$, each class
being the $GL(n)$ orbit $\calO(P)$ of any point $P$ in the class. 
The crucial observation to be made here is that 
there exist two
types of points on $\calL_n$: those that are completely surrounded
by equivalent points and those whose neighborhoods
include non-equivalent points, sketched as $P$ and $Q$ respectively in
Fig.~1. Any infinitesimal perturbation of the structure
constants of $\calG_P$ will
necessarily lead to an isomorphic Lie algebra while there exist
infinitesimal perturbations of $\calG_Q$ that lead outside of
$\calO(Q)$ and, hence, to non-isomorphic algebras. 

Given a Lie algebra $\calG$ with the Lie product of $T_A$, $T_B \in
\calG$ supplied by
the commutator $[T_A, \, T_B]_0$. A {\em one-parameter deformation} 
of $\calG$ is given by the {\em deformed commutator}
\ble{defcom}
[T_A,\ T_B]_t = [T_A, \, T_B]_0 
                + \sum_{m=1}^\infty \psi^{(2)}_m(T_A, \, T_B) \, t^m
\, ,
\ee
where $t$ is a formal parameter and the $\psi^{(2)}_m$ are 
$\calG$-valued, bilinear antisymmetric maps 
\ble{psim}
\psi^{(2)}_m: \quad \calG \times \calG \rightarrow \calG
\, ,
\qquad \qquad 
\psi^{(2)}_m(T_A, \, T_B) = - \psi^{(2)}_m(T_B, \, T_A) 
\, .
\ee
Such maps are called {\em 2-cochains} --- in a similar fashion
one defines $p$-{\em cochains} $\psi^{(p)}$ which accept $p$ arguments. 
The {\em
coboundary operator} $s$ maps $p$-cochains to $p+1$-cochains
according to
\bae
\label{cobdef}
s \triangleright \psi^{(p)} (T_{A_1}, \ldots ,T_{A_{p+1}}) 
\fe
\sum_{r=1}^{p+1} (-1)^{p+1} [T_{A_r}, \, \psi^{(p)} (T_{A_1},
\ldots ,\hat{T}_{A_r}, \ldots ,T_{A_{p+1}} )]
\ff
 & & + \sum_{r<s} (-1)^{r+s} \psi^{(p)} ([T_{A_r},
\, T_{A_s}], \, T_{A_1}, \ldots ,\hat{T}_{A_r}, \ldots
,\hat{T}_{A_s},
\ldots , T_{A_{p+1}} )
\, .
\eae
We will have more to say about
$s$ later on. For the moment, we just state that $s^2=0$. Then, 
mimicking
a familiar procedure in differential geometry, one defines
$p$-{\em cocycles} as $p$-cochains annihilated by $s$ (spanning
$Z^p$), $p$-{\em
coboundaries} (or, {\em trivial} $p$-cocycles) as the image, under 
$s$, of $(p-1)$-cochains (spanning $B^p$) and the
$p$-{\em th cohomology group} of $\calG$ as $Z^p/B^p \equiv H^p$. The
relevance of all this technology to Lie algebra deformations
can be seen by differentiating, w.r.t. $t$, the Jacobi identity that 
$[\cdot, \, \cdot]_t$ must satisfy and then setting $t=0$. One
finds that $\psi^{(2)}_1$ in~(\ref{defcom}) has to be a 2-cocycle. 
Moreover, if $\psi^{(2)}_1$ is a coboundary, the deformation that it
generates is trivial, in the sense that the resulting deformed Lie 
algebra is isomorphic to the original one. Referring to
Fig.~1, we see that 2-cocycles span the tangent space
to $\calL_n$ at, \eg, $Q$, with 2-coboundaries pointing towards the
$GL(n)$ orbit of $Q$ and non-trivial 2-cocycles outside of it.
A sufficient condition then for the stability of $\calG$ is the
vanishing of its second cohomology group $H^2(\calG)$.
It is worth pointing out that the above is not a necessary
condition. Although a non-trivial 2-cocycle may exist,
obstructions originating in $H^3(\calG)$ can render it
non-integrable in which case the corresponding finite 
non-trivial deformation does not exist (see, \eg,~\cite{Ric:67}).

It is obvious from the definition given above, that a $p$-cochain
can always be realized as a $\calG$-valued left invariant (LI) 
$p$-form on the group manifold $G$ corresponding to $\calG$, with the
generators $T_A$ now extended to LI vector fields. Denoting by
$\Pi^A$  the LI 1-forms on $G$, we write $\psi^{(p)}$ as
\ble{psiLI}
\psi^{(p)} \equiv T_B \otimes \psi^B = \frac{1}{p!} 
              {\psi_{A_1 \ldots A_p}}^B \,T_B \otimes \Pi^{A_1}
              \ldots \Pi^{A_p}
\, .
\ee
Then the action of $s$ given in~(\ref{cobdef}) coincides with
that of a covariant exterior derivative $\nabla$, 
\ble{nabladef}
\nabla (T_A \otimes \psi^A) = T_A \otimes (d \psi^A +
{\Omega^A}_B \psi^B)
\, ,
\ee
with the
connection 1-form $\Omega$ given by 
\ble{conndef}
{\Omega^A}_B = {f_{RB}}^A \Pi^R
\, ,
\qquad
(\mbox{\ie,}
\quad
\nabla_{T_A} T_B = [T_A, \, T_B])
\, .
\ee
The nilpotency of $s$ follows now from the vanishing of the 
curvature 2-form
$\Theta = d\Omega + \Omega^2$, due to the Jacobi identity, while 
2-cocycles are covariantly constant $\calG$-valued LI
2-forms. Notice that the requirement that $s \triangleright 
\psi^{(2)} = 0$, with
$\psi^{(2)}$ as in~(\ref{psiLI}), reduces to 
\ble{JIlin}
  {f_{AR}}^S {\psi_{BC}}^R 
+ {f_{BR}}^S {\psi_{CA}}^R 
+ {f_{CR}}^S {\psi_{AB}}^R 
+ {\psi_{AR}}^S {f_{BC}}^R 
+ {\psi_{BR}}^S {f_{CA}}^R 
+ {\psi_{CR}}^S {f_{AB}}^R 
= 0
\, ,
\ee
which is, as expected, the linearized form of the Jacobi identity.

Imagine one is given all (non-zero) commutators 
that define a Lie algebra $\calG$ and  is asked to strategically 
add to their r.h.s. (multiples of) a new, central generator $Z$,
without violating the Jacobi identity. Assuming that this is at all
possible, the resulting Lie algebra is called a {\em central
extension} of $\calG$. It is easily seen that central extensions
can be considered as a particular class of deformations --- 
one simply pretends
that $Z$ was present all the time as a $U(1)$ factor, and got
involved in the algebra, remaining central, as a result of the 
deformation. The general results above simplify considerably in
this case. The maps $\psi^{(2)}_m$ in~(\ref{defcom}) are valued
now in the center of $\calG$ and, as a result, the first sum on 
the r.h.s. of~(\ref{cobdef}) vanishes.
What is left reduces, in our geometrical realization of $s$,
to the exterior derivative $d$. 2-cocycles become, in this case,
closed LI 2-forms, 2-coboundaries are exact LI 2-forms {\em with
LI potential} while 
non-trivial 2-cocycles are closed LI 2-forms that do not have LI
potential 1-form. The latter are known as {\em non-trivial 
Chevalley-Eilenberg (CE) 2-cocycles}, with obvious generalization
to $p$-cocycles. Notice that a non-trivial
CE cocycle {\em may} admit a potential, although not a LI one.

We wrap up our brief tour of things deformed with a couple of
remarks.
First, the discussion of stability above implies that the
dimensions of $\calL_n$ and $\calO(P)$ must be equal in order for
$\calG_P$ to be stable (see
Fig.~1) --- this
would seem to single out a couple of special values of $n$ for
which stability is at all possible. However, in physical
applications, one generally has to take into account additional
restrictions on the structure constants (\eg, Lorentz covariance)
which make the balancing of dimensions more complicated and give
rise to several possible values of $n$.
Furthermore, apart from the mandatory physical restrictions, one
may elect to consider only certain types of deformations, leading
to a corresponding (weaker) notion of partial stability\footnote{%
we thank Gary Gibbons for discussions on this point.%
}. Second, we have dealt above, for the sake of simplicity,  
with ordinary Lie algebras --- analogous results hold for Lie
superalgebras~\cite{Nij.Ric:66}, which are the ones we deal 
with in later sections.

As an example of a stabilizing deformation, consider the
(homogeneous) Galilean algebra
\ble{Gal}
[J_i, \, J_j]=\epsilon_{ijk} J_k
\, ,
\qquad \qquad
[J_i, \, K_j]=\epsilon_{ijk} K_k
\, ,
\qquad \qquad
[K_i, \, K_j]=0
\, .
\ee
Denoting by $\Pi^i$, $\Pi^{\bar{i}}$ the LI 1-forms corresponding 
to $J_i$, $K_i$ respectively, one easily shows that
\ble{psiG}
\psi^{(2)}_1 = 
\frac{1}{2} J_1 \otimes \Pi^{\bar{2}} \Pi^{\bar{3}}
+ \frac{1}{2} J_2 \otimes \Pi^{\bar{3}} \Pi^{\bar{1}}
+ \frac{1}{2} J_3 \otimes \Pi^{\bar{1}} \Pi^{\bar{2}}
\ee
is a non-trivial 2-cocycle (notice that the $\calG$-part of
$\psi^{(2)}_1$ above does not commute with $\calG$ and, hence, the full
expression~(\ref{nabladef}) has to be used, with $d
\Pi^{\bar{1}}= -\Pi^{2} \Pi^{\bar{3}} + \Pi^3 \Pi^{\bar{2}}$
\etc.). The resulting 
deformation is
\ble{Galdef}
[J_i, \, J_j]=\epsilon_{ijk} J_k
\, , 
\qquad \qquad
[J_i, \, K_j]=\epsilon_{ijk} K_k
\, ,
\qquad \qquad
[K_i, \, K_j]=-\frac{1}{c^2} \epsilon_{ijk} J_k
\, ,
\ee
where the scale of $\psi^{(2)}_1$, undetermined by the cocycle
condition, is fixed by the experiment. This is the Lorentz
algebra and, being semisimple, is stable, according to
Whitehead's second lemma~\cite{Jac:62}.

A second, substantially more involved, example 
is furnished by the passage from classical to quantum
mechanics. The complexity in this case arises from the fact that the 
setting amenable to stability analysis is that of the 
infinite-dimensional 
Lie algebra of functions $f(q,p)$ on phase space, with the Lie product
given by the Poisson bracket --- for details,
see~\cite{Bay.Fla.Fro.Lic.Ste:78}, while an alternative,
finite-dimensional treatment is given in~\cite{Vil:94}. 
%%%%%%%%%%%%%%%%%%%%%%%%%%%%%%%%%%%%%%%%%%%%%%%%%%%%%%%%%%%%%%%%%
%%%%%%%%%%%%%%%%%%%%%%%%%%%%%%%%%%%%%%%%%%%%%%%%%%%%%%%%%%%%%%%%%
\section{Quasi-invariant Actions} 
\label{QiA}
%%%%%%%%%%%%%%%%%%%%%%%%%%%%%%%%%%%%%%%%%%%%%%%%%%%%%%%%%%%%%%%%%
%%%%%%%%%%%%%%%%%%%%%%%%%%%%%%%%%%%%%%%%%%%%%%%%%%%%%%%%%%%%%%%%%
We lay coordinates $\xi^i, \,
i=0, \dots,p $ on the $(p+1)$-dimensional worldvolume $W$ of a
$p$-brane and embed it in $D$-dimensional Minkowski spacetime $M$,
with coordinates $x^\mu, \, \mu=0, \ldots,D-1$. The embedding,
which we will generically denote by $\phi$, is effected by functions
$x^\mu(\xi)$. The action is taken to be the hypervolume swept
out by the brane, with metric induced by the Minkowskian
$\eta_{\mu \nu}$
\ble{braction}
S_p = - \int_W d^{p+1} \xi \, \sqrt{-\det m_{ij}}
\, ,
\qquad \qquad
m_{ij} = \frac{\partial x^\mu}{\partial \xi^i} 
\frac{\partial x^\nu}{\partial \xi^j} \eta_{\mu \nu}
\, .
\ee 
$M$ is a group manifold, with the group operation being ordinary 
translations, ${x''}^\mu= {x'}^\mu + x^\mu$, the generators
$X_\mu$ of which obviously commute. These leave invariant 
the above action --- in fact, they even leave invariant the Lagrangian
density, \ie, the action is invariant even
when surface terms cannot be ignored. 

We now wish to consider the motion of our brane in superspace
$\Sigma$. The latter is just like  $M$, but with additional
fermionic coordinates $\theta^\alpha$, the embedding $\phi$ now given
by  functions $\theta^\alpha(\xi)$, $x^\mu(\xi)$. $\Sigma$ is
also a (super\footnote{%
we use the prefix {\em super} only minimally.%
}%
)group, the group operation being {\em
supertranslations}
\ble{strans}
\theta''^\alpha = \theta'^\alpha + \theta^\alpha
\, ,
\qquad \qquad \qquad
x''^\mu = x'^\mu +x^\mu +{1 \over 2} (C \Gamma^\mu)_{\alpha\beta}
            \theta'^\alpha \theta^\beta
\, .
\ee
$C$ here is the charge conjugation matrix, $C
\Gamma_\mu C^{-1} = -\Gamma_\mu^T$.
The corresponding generators $D_\alpha$, $X_\mu$ satisfy
\ble{salg}
\{ D_\alpha, D_\beta\} = \GMab X_\mu
\, ,
\ee
all other commutators being zero. At this point we can establish 
a connection with the material of the previous section:
(\ref{salg}) is clearly a central extension of the commutative
fermionic algebra $\{ D_\alpha, D_\beta\} = 0$, by $X_\mu$.
Notice how this observation points to the fermionic sector of
$\Sigma$ as the fundamental one, $M$ entering only as a stabilizer.
Rewriting~(\ref{salg}) in the dual form
\ble{salgMC}
d \Pi^\alpha = 0
\, ,
\qquad \qquad \qquad
d \Pi^\mu = {1 \over 2} \GMab \Pi^\alpha \Pi^\beta
\, ,
\ee
we identify $\GMab \Pi^\alpha \Pi^\beta$ as the
non-trivial CE 2-cocycle responsible for the extension.
$\Pi^\alpha$, $\Pi^\mu$ here are the LI 1-forms on $\Sigma$,
given by
\ble{Piam}
\Pi^\alpha = d \theta^\alpha
\, ,
\qquad \qquad \qquad
\Pi^\mu = d x^\mu + {1\over 2} \GMab \theta^\alpha d\theta^\beta
\, .
\ee
Notice that left invariance simply means invariance under (left)
supertranslations.
For this reason, the $\Pi$'s are the natural building blocks for a 
supertranslation invariant action, which takes now the form
\ble{brsaction}
S_p = - \int_W d^{p+1} \xi \, \sqrt{-\det m_{ij}}
\, ,
\qquad \qquad
m_{ij} = \Pi^\mu_i \Pi^\nu_j  \eta_{\mu \nu}
\, ,
\ee 
where $\phi^*(\Pi^\mu) \equiv \Pi^\mu_i d\xi^i$ is the pull-back of
$\Pi^\mu$ on $W$. We are not yet finished though. As mentioned 
already in Sec.{}~\ref{Intro}, $\kappa$-symmetry forces the addition of
a Wess-Zumino term $S_{WZ}$ to the above action, given by the
pull-back on $W$ of a $(p+1)$-form $b$ on $\Sigma$,
\ble{Stotdef}
S = S_p  + S_{WZ}
\, ,
\qquad \qquad
S_{WZ} = \int_W \phi^*(b)
\, .
\ee
The properties required of $b$ are best expressed in terms of $h =
db$: $h$ must be a closed LI and Lorentz invariant $(p+2)$-form
with length dimension $p+1$. The latter is computed by assigning
length dimension 1 to $\Pi^\mu$ and $\frac{1}{2}$ to $\Pi^\alpha$.   
One finds
\ble{hdef}
h = (C \Gamma_{\mu_1\dots \mu_p})_{\alpha\beta} 
\Pi^{\mu_1}\dots \Pi^{\mu_p} \Pi^\alpha \Pi^\beta
\, ,
\ee
where $\Gamma_{\mu_1\dots \mu_p}$ is the antisymmetrized product
(with unit weight) of $\Gamma_{\mu_1}, \dots ,\Gamma_{\mu_p}$.

The natural question that arises is whether $h$ comes  
from a LI potential $b$, so that the invariance of the total
action $S$ is maintained. The answer is a categorical
no~\cite{Azc.Tow:89}, which promotes $h$ to a non-trivial
CE $(p+2)$-cocycle. $h$ does come from a $(p+1)$-form $b$, but
the latter transforms by a total derivative under
supertranslations, making $S$ only quasi-invariant. 
Siegel has showed that by enlarging
superspace by a new fermionic variable, a genuinely invariant
action can be found in the $p=1$ case
(superstring)~\cite{Sie:94}. At the algebra level, this is
just a central extension of $\Sigma$. The extended
algebra does not admit any further extension by generators with a
single Lorentz index, vectorial or spinorial, and it is only this type
of generators that are relevant in the description of the superstring.
We conclude that Siegel's extension of $\Sigma$ is a
stabilizing deformation, as long as one only admits generators
with a single Lorentz index (this is an example of the partial
stability mentioned in Sec.{}~\ref{Intro}). In the case of 
$p$-branes then, it
is only natural to ask the question: what is the stable form of
the supersymmetry algebra~(\ref{salg}) (or~(\ref{salgMC})), when 
only additional generators with $p$ Lorentz indices are allowed?  
%%%%%%%%%%%%%%%%%%%%%%%%%%%%%%%%%%%%%%%%%%%%%%%%%%%%%%%%%%%%%%%%%
%%%%%%%%%%%%%%%%%%%%%%%%%%%%%%%%%%%%%%%%%%%%%%%%%%%%%%%%%%%%%%%%%
\section{Extensions} 
\label{Ext}
%%%%%%%%%%%%%%%%%%%%%%%%%%%%%%%%%%%%%%%%%%%%%%%%%%%%%%%%%%%%%%%%%
%%%%%%%%%%%%%%%%%%%%%%%%%%%%%%%%%%%%%%%%%%%%%%%%%%%%%%%%%%%%%%%%%
Our starting point is the supersymmetry algebra in its
Maurer-Cartan (MC) form, Eq.{}~(\ref{salgMC}), with $\Pi^\mu$
rescaled so as to keep the
discussion as general as possible
\ble{salgres}
d \Pi^\alpha =0
\, ,
\qquad \qquad
d \Pi^\mu = a_s \GMab \Pi^\alpha \Pi^\beta
\, .
\ee
We look for
central extensions involving generators with $p$ Lorentz
indices, \ie, for non-trivial CE 2-cocycles. We find that 
\ble{rho10}
\rho_{\mu_1 \dots \mu_p} = 
(C \Gamma_{\mu_1 \dots \mu_p})_{\alpha\beta}
\Pi^\alpha \Pi^\beta
\ee
is the only such cocycle available. We introduce accordingly a new
LI 1-form $\Pi_{\mu_1 \dots \mu_p}$ and set its differential
equal to the cocycle,
\ble{Pirho}
d \Pi_{\mu_1 \dots \mu_p} = 
a_0 (C \Gamma_{\mu_1 \dots \mu_p})_{\alpha\beta}
\Pi^\alpha \Pi^\beta
\, .
\ee
On the group manifold, this amounts to the introduction of a new
coordinate\footnote{%
what we mean is of course a new {\em type} of coordinate ---
there are ${D \choose p}$ of them.%
}
$\phi_{\mu_1 \dots \mu_p}$, so that the set $\{\theta^\alpha,\,
x^\mu, \, \phi_{\mu_1 \dots \mu_p} \}$ parameterizes now the {\em
extended} superspace $\tilde{\Sigma}_0$. The relation of
$\phi_{\mu_1 \dots \mu_p}$ to $\Pi_{\mu_1 \dots
\mu_p}$ as well as its group law are uniquely determined by the
non-LI potential for $\rho_{\mu_1 \dots \mu_p}$,
see~\cite{Chr.Azc.Izq.Bue:00} for details. The only non-zero
commutator of the resulting Lie algebra is
\ble{LA1ext}
\{ D_\alpha, D_\beta\} = a_s \GMab X_\mu + 
a_0 (C \Gamma_{\mu_1 \dots \mu_p})_{\alpha\beta}
Z^{\mu_1 \dots \mu_p}
\, ,
\ee
where $Z^{\mu_1 \dots \mu_p}$ denotes the generator of
translations along $\phi_{\mu_1 \dots \mu_p}$. We now repeat the
above procedure with $\tilde{\Sigma}_0$ replacing $\Sigma$. We
find that the introduction of $\Pi_{\mu_1 \dots \mu_p}$, which
repaired the instability of the original algebra, has as a
side effect the appearance of a new non-trivial CE 2-cocycle on
the extended superspace $\tilde{\Sigma}_0$. Indeed, inspection of
the available Lorentz tensors shows that the only two
candidates for a cocycle are\footnote{%
we use the index notation $\mu_1 \dots \alpha_k \equiv \mu_1 \dots
\mu_{p-k} \alpha_1 \dots \alpha_k$.%
}
\ble{cand1}
\rho^{(1)}_{\mu_1 \dots \alpha_1}=(C \Gamma_{\nu \mu_1
\dots
\mu_{p-1}})_{\beta \alpha_1} \Pi^\nu \Pi^\beta
\, ,
\qquad 
\qquad 
\qquad 
\rho^{(2)}_{\mu_1 \dots \alpha_1}= 
(C \Gamma^\nu)_{\beta \alpha_1} 
\Pi_{\nu \mu_1 \dots \mu_{p-1}} \Pi^\beta
\, . 
\ee
For $p=1$, both of these forms are closed. For $p \geq 2$, 
$\lambda_2={a_s \over a_0}$ implies $d (\rho^{(1)} + 
\lambda_2 \rho^{(2)})=0$ 
provided\footnote{%
primed indices are understood to be symmetrized with unit
weight.%
}
\ble{Gammaprop}
(C \Gamma^\nu)_{\alpha' \beta'} (C \Gamma_{\nu \mu_1 \dots
\mu_{p-1}})_{\gamma' \delta'}=0
\, ,
\ee
a relation which is true only for certain values of
$(D,p)$~\cite{Ach.Eva.Tow.Wil:87}. We introduce then a LI 1-form
$\Pi_{\mu_1 \dots \alpha_1}$ and the corresponding coordinate
$\phi_{\mu_1 \dots \alpha_1}$, obtaining the new extended superspace
$\tilde{\Sigma}_1$\footnote{%
notice that the subscript of $\tilde{\Sigma}$ counts the number
of fermionic indices of the last coordinate introduced.%
}. 
The MC equations are augmented by
\ble{dPim1a1}
d\Pi_{\mu_1 \dots \alpha_1} = 
a_1 
\left(
  (C \Gamma_{\nu
  \mu_1 \dots \mu_{p-1}})_{\beta \alpha_1} \Pi^\nu \Pi^\beta
  + {a_s \over a_0} (C\Gamma^\nu)_{\beta \alpha_1} \Pi_{\nu \mu_1
  \dots \mu_{p-1}} \Pi^\beta)
\right)
\, .
\ee
Notice that this new extension
involves the replacement of a vectorial index by a spinorial one.
Also, the appearance of $\rho^{(2)}_{\mu_1 \dots \alpha_1}$
in $d\Pi_{\mu_1 \dots \alpha_1}$, implies that $Z^{\mu_1 
\dots \mu_p}$ is no
longer central. Analogous  remarks hold for all subsequent extensions.
Every new generator introduced has one more spinorial index than
the previous one, while, after each extension, the only central
generator is the one introduced last, all others 
having acquired non-zero commutators as a result of the extensions
made after the one that introduced them. The iterative procedure
stops with the introduction of $Z_{\alpha_1 \dots \alpha_p}$,
since there are no non-trivial 2-cocycles on $\tilde{\Sigma}_p$. 
The first five
extensions are, from an algebraic point of view, exceptional,
while for the ones that follow a pattern emerges that leads to a
recursion relation. The resulting MC equations are (apart
from~(\ref{salgres}), (\ref{Pirho}), (\ref{dPim1a1}))  
\begin{eqnarray}
\label{dPim1ak2}
%%%%%%%%%%%%%%%%%%%%%%%%%%%%
d \Pi_{\mu_1 \dots \alpha_2} \fe
a_2 \left\{ (C \Gamma_{\nu \rho \mu_1 \dots \mu_{p-2}})_{ \alpha_1
\alpha_2} \Pi^\nu \Pi^\rho
+ {a_s \over a_0} (C \Gamma^\nu)_{\alpha_1 \alpha_2}
  \Pi_{\nu \rho \mu_1 \dots \mu_{p-2}} \Pi^\rho \right. \nonumber \\
& & - {a_s \over a_1}  \left. (C \Gamma^\nu)_{\alpha_1 \alpha_2}
   \Pi_{\nu \mu_1 \dots \mu_{p-2} \beta} \Pi^\beta
   -8 {a_s \over a_1}  (C \Gamma^\nu)_{\alpha'_1 \beta}
   \Pi_{\nu \mu_1 \dots \mu_{p-2} \alpha'_2} \Pi^\beta \right\}
\, ,
\ff
%%%%%%%%%%%%%%%%%%%%%%%%%%%%
d \Pi_{\mu_1 \dots \alpha_3} 
\fe
  a_3 
  \left\{ 
  (C \Gamma^\nu)_{\alpha'_1 \alpha'_2}
  \Pi_{\nu \rho \mu_1 \dots \mu_{p-3} \alpha'_3} \Pi^\rho 
  + {5a_1 \over 4a_2}  (C \Gamma^\nu)_{\alpha'_1 \beta}
  \Pi_{\nu \mu_1 \dots \mu_{p-3} \alpha'_2 \alpha'_3} \Pi^\beta
  \right.
\ff
& & 
+ {a_1 \over 4a_2}  \left. (C \Gamma^\nu)_{\alpha'_1 \alpha'_2}
   \Pi_{\nu \mu_1 \dots \mu_{p-3} \beta \alpha'_3} \Pi^\beta
  \rule[-3.5mm]{0mm}{7mm}
\right\}
\, ,
\\
%%%%%%%%%%%%%%%%%%%%%%%%%%%%
d \Pi_{\mu_1 \dots \alpha_4} 
\fe
  a_4 
  \left\{ 
    (C \Gamma^\nu)_{\alpha'_1 \alpha'_2}
    \Pi_{\nu \rho \mu_1 \dots \mu_{p-4} \alpha'_3 \alpha'_4} \Pi^\rho
    - {48a_sa_2 \over 5a_1 a_3}  (C \Gamma^\nu)_{\alpha'_1 \beta}
    \Pi_{\nu \mu_1 \dots \mu_{p-4} \alpha'_2 \alpha'_3 \alpha'_4}
    \Pi^\beta 
  \right.
\ff
& & 
  - {12a_s a_2\over 5a_1 a_3}  \left. 
  (C \Gamma^\nu)_{\alpha'_1 \alpha'_2}
  \Pi_{\nu \mu_1 \dots \mu_{p-4} \beta \alpha'_3 \alpha'_4} \Pi^\beta
  \rule[-3.5mm]{0mm}{7mm}
\right\}
\, ,
\ff
%%%%%%%%%%%%%%%%%%%%%%%%%%%%
d \Pi_{\mu_1 \dots \alpha_{k+2}} 
\fe
a_{k+2} \left\{ (C \Gamma^\nu)_{\alpha'_1 \alpha'_2}
  \Pi_{\nu \rho \mu_1 \dots \mu_{p-(k+2)} \alpha'_3 \dots
   \alpha'_{k+2}} \Pi^\rho  
    + \lambda^{(k+2)}_2 (C \Gamma^\nu)_{\alpha'_1 \beta}
   \Pi_{\nu \mu_1 \dots \mu_{p-(k+2)} \alpha'_2 \dots \alpha'_{k+2}}
   \Pi^\beta 
\right.
\ff
& & 
   \left. + \lambda^{(k+2)}_3 (C \Gamma^\nu)_{\alpha'_1 \alpha'_2}
   \Pi_{\nu \mu_1 \dots \mu_{p-(k+2)} \beta \alpha'_3 \dots
   \alpha'_{k+2}}
   \Pi^\beta \right\}
\, ,
\qquad k=3,4,\dots
\, ,
\nonumber
%%%%%%%%%%%%%%%%%%%%%%%%%%%%
\end{eqnarray}
where
\ble{la23}
\lambda^{(k+2)}_2  =  -{a_s \over a_{k+1}}\left( {2 \over
\lambda^{(k+1)}_2} + {k \over \lambda^{(k+1)}_3}\right) 
\, , 
\qquad \qquad
\lambda^{(k+2)}_3  =  -{a_s \over a_{k+1}}{k+1 \over
\lambda^{(k+1)}_2}
\, .
\ee
$\tilde{\Sigma}_p$ has a fibre
bundle structure, with $\Sigma$ in the base and the new
coordinates $\{\phi_{\mu_1 \dots \mu_p},\dots , \phi_{\alpha_1
\dots \alpha_p} \}$ along the fiber. Denoting symbolically by $D$,
$Y$ the corresponding (classes of) generators,
the Lie algebra dual to the above MC equations takes the form
\ble{LAsymb}
[D,D] \sim D+Y
\, , \qquad \qquad \quad
[D,\, Y] \sim Y
\, , \qquad \qquad \quad
[Y,\, Y] = 0
\, .
\ee
Although our way of constructing the above algebra guarantees
its stability against central extensions by generators with
$p$ Lorentz indices, it is not {\em a priori} clear that it is
stable under general deformations  involving the above
generators. In other words, although we know that there are no
non-trivial CE 2-cocycles on $\tilde{\Sigma}_p$, we have not
explicitly shown that there are no non-trivial 2-cocycles for the
full coboundary operator $s$, Eq.~(\ref{cobdef}). That this is
indeed the case can actually be inferred from the results
of~\cite{Sez:97}. We conclude that {\em the Lie superalgebra
given by~(\ref{salgres}), (\ref{Pirho}), (\ref{dPim1a1}),
(\ref{dPim1ak2}), (\ref{la23}) is stable under deformations
involving generators with $p$ Lorentz indices}.
%%%%%%%%%%%%%%%%%%%%%%%%%%%%%%%%%%%%%%%%%%%%%%%%%%%%%%%%%%%%%%%%%
%%%%%%%%%%%%%%%%%%%%%%%%%%%%%%%%%%%%%%%%%%%%%%%%%%%%%%%%%%%%%%%%%
\section{Applications} 
\label{Appli}
%%%%%%%%%%%%%%%%%%%%%%%%%%%%%%%%%%%%%%%%%%%%%%%%%%%%%%%%%%%%%%%%%
%%%%%%%%%%%%%%%%%%%%%%%%%%%%%%%%%%%%%%%%%%%%%%%%%%%%%%%%%%%%%%%%%
%%%%%%%%%%%%%%%%%%%%%%%%%%%%%%%%%%%%%%%%%%%%%%%%%%%%%%%%%%%%%%%%%
\subsection{Invariant actions} 
\label{Actions}
%%%%%%%%%%%%%%%%%%%%%%%%%%%%%%%%%%%%%%%%%%%%%%%%%%%%%%%%%%%%%%%%%
We may now return to the question of finding a LI potential
$\tilde{b}$ for the 
$h$ in Eq.~(\ref{hdef}). In $\tilde{\Sigma}_p$ this is indeed
possible --- it takes the general form~\cite{Chr.Azc.Izq.Bue:00}
\ble{tbdef}
\tilde{b} = \sum_{k=0}^p  b_k \, \Pi_{\mu_1 \dots \alpha_k} 
   \Pi^{\mu_1} \dots  \Pi^{\alpha_k}
\, ,
\ee
where the $b_k$ are numerical constants, determined by $h=d\tilde{b}$. 
For example, for $D=10$, $p=1$ we find Siegel's result\footnote{%
notice that we never use the metric to raise or lower indices, so
that the position of the Lorentz indices can serve to identify
the generators, forms \etc.. For example, $\Pi_\mu$
in~(\ref{bp1}) is the LI 1-form corresponding to the new
coordinate $\phi_\mu$ and has no relation to $\Pi^\mu$ (similarly
for $\Pi_\alpha$).%
}
\ble{bp1}
\tilde{b} = \Pi_\mu \Pi^\mu + \frac{1}{2} \Pi_\alpha \Pi^\alpha
\, ,
\ee
while, for $D=11$, $p=2$, we get
\ble{bp2}
\tilde{b}={2 \over 3} \Pi_{\mu \nu} \Pi^\mu \Pi^\nu
- {3 \over 5} \Pi_{\mu \alpha} \Pi^\mu \Pi^\alpha
- {2 \over 15} \Pi_{\alpha \beta} \Pi^\alpha \Pi^\beta
\, ,
\ee
in accordance with~\cite{Ber.Sez:95}. Explicit expressions for
the above two cases, including the extended Lie algebras, MC
equations, group law, LI vector fields and Noether currents,
can be found in~\cite{Chr.Azc.Izq.Bue:00}. 
%%%%%%%%%%%%%%%%%%%%%%%%%%%%%%%%%%%%%%%%%%%%%%%%%%%%%%%%%%%%%%%%%
\subsection{\mathversion{bold} $D$-branes \mathversion{normal}} 
\label{Db}
%%%%%%%%%%%%%%%%%%%%%%%%%%%%%%%%%%%%%%%%%%%%%%%%%%%%%%%%%%%%%%%%%
We  sketch the relevance of $\tilde{\Sigma}_p$ in the
description of $Dp$-branes, using as example the $D=10$, $p=2$
IIA case. 
The qualitative novelty here, compared to standard $p$-branes,  
is the appearance of the Born - Infeld field $A_i(\xi)$, 
defined directly on
the worldvolume. Our starting point will be the abstract 
free differential algebra (FDA)\footnote{%
The term refers to an algebra 
generated by differential forms which is closed under the action
of $d$, in such a way as to have $d^2=0$.%
}
\bae
\label{D2fda}
d\Pi^\mu 
\fe 
\frac{1}{2}(C\Gamma^\mu)_{\alpha\beta} \Pi^\alpha\Pi^\beta
\, ,
\ff
d\Pi^\alpha \fe 0
\, , 
\ff
d{\cal F} 
\fe 
\Pi^\mu(C\Gamma_\mu\Gamma_{11})_{\alpha\beta} \Pi^\alpha\Pi^\beta
\, ,
\eae
where ${\cal F}\equiv dA-B$ is invariant and 
\ble{dB}
dB=-(C\Gamma_\mu\Gamma_{11})_{\alpha\beta}\Pi^\mu \Pi^\alpha\Pi^\beta
\, .
\ee
For the above FDA $d^2=0$ since, in $D=10$ we have
$(C\Gamma^\mu\Gamma_{11})_{\alpha{'}\beta{'}}
(C\Gamma_\mu)_{\delta{'}\epsilon{'}}=0$. One looks for non-trivial
$(p+2)$-cocycles $h$, constructed from the above forms, with
length dimension $p+1$, so as to match that of the kinetic
Lagrangian. Such cocycles exist only for $p \leq 8$ and even,
\ie, precisely for those values for which $Dp$-branes of type IIA
are known to exist. For $p=2$, $h$ takes the form
\ble{hDp2}
h = (C\Gamma_{\mu\nu})_{\alpha\beta}
    \Pi^\mu \Pi^\nu \Pi^\alpha \Pi^\beta
    -(C\Gamma_{11})_{\alpha\beta}
    \Pi^\alpha \Pi^\beta {\cal F}
\, ,
\ee
while its LI potential follows (apart from the $\calF$ term) from
an appropriate dimensional reduction of~(\ref{bp2}),
\ble{bDp2}
{\tilde b}=
 \frac{2}{3} \Pi_{\mu\nu} \Pi^\mu \Pi^\nu  
 + \frac{4}{3} \Pi_\mu \Pi^\mu \Pi 
 - \frac{2}{15} \Pi_{\alpha\beta} \Pi^\alpha \Pi^\beta
 -\frac{3}{5} \Pi_{\mu\alpha} \Pi^\mu \Pi^\alpha  
 - \frac{3}{5} \Pi_\alpha \Pi \Pi^\alpha 
 - 2 \Pi \calF
\, .
\ee
Moreover, 
\ble{dPP}
d \bigl( 
\frac{1}{2} \Pi^\alpha \Pi_\alpha - \Pi^\mu \Pi_\mu
\bigr)  
 = (C \Gamma_\mu \Gamma_{11})_{\alpha \beta} 
   \Pi^\mu \Pi^\alpha \Pi^\beta
\, ,
\ee
which shows that one may set, on the extended superspace,
\ble{Fext}
\calF =
\frac{1}{2} \Pi^\alpha \Pi_\alpha - \Pi^\mu \Pi_\mu
\, .
\ee
Finally, the worldvolume field $A$ may be identified with the
pull-back, on $W$, of the following 1-form
\ble{Aext}
A=\varphi_\mu dx^\mu + \frac{1}{2} \varphi_\alpha d \theta^\alpha
\, ,
\ee
defined also on the extended superspace. Similar results hold for
the $D=11$ $M5$-brane --- the details may be found 
in~\cite{Chr.Azc.Izq.Bue:00}.
%%%%%%%%%%%%%%%%%%%%%%%%%%%%%%%%%%%%%%%%%%%%%%%%%%%%%%%%%%%%%%%%
%%%%%%%%%%%%%%%%%%%%%%%%%%%%%%%%%%%%%%%%%%%%%%%%%%%%%%%%%%%%%%%%
\section{Epilogue} 
\label{Epi}
%%%%%%%%%%%%%%%%%%%%%%%%%%%%%%%%%%%%%%%%%%%%%%%%%%%%%%%%%%%%%%%%%
%%%%%%%%%%%%%%%%%%%%%%%%%%%%%%%%%%%%%%%%%%%%%%%%%%%%%%%%%%%%%%%%%
Our aim has been to argue that Lie (super)algebra stability is a
sensible criterion for the soundness of a physical theory. To the
standard list of examples, we may now add the results
of~\cite{Chr.Azc.Izq.Bue:00}, where the natural emergence of
extended superspaces in the description of extended objects
leads to  
the resolution of a number of problems arising in the
conventional formulation. Assuming a (even partially) convinced 
reader, we take a
further step in this direction  proposing that the recent
proliferation of non-commutative spaces, as the arenas for
physical phenomena, should be examined in the light of stability
considerations. In particular, the form of the spacetime
coordinate commutation relations could be inferred, or at least
narrowed down, by demanding the stability, under suitable
restrictions, of the resulting Lie algebra, much in the spirit of
the Poincar\'e algebra analysis in~\cite{Vil:94}. 
The novelty that arises, in view of our results, is the
possibility to include 
the additional coordinates of the extended spaces
encountered.  We plan on
pursuing this theme in a forthcoming publication~\cite{Chr:01b}.
%%%%%%%%%%%%%%%%%%%%%%%%%%%%%%%%%%%%%%%%%%%%%%%%%%%%%%%%%%%%%%%%%
%%%%%%%%%%%%%%%%%%%%%%%%%%%%%%%%%%%%%%%%%%%%%%%%%%%%%%%%%%%%%%%%%
\section*{Acknowledgements}
%%%%%%%%%%%%%%%%%%%%%%%%%%%%%%%%%%%%%%%%%%%%%%%%%%%%%%%%%%%%%%%%%
Warm thanks are due to the organizers of the Torino Conference 
``Brane New
World and Non-commutative Geometry'' for all their efforts, also
for financial support. It was a memorable week, full of
non-commutative excitement, high above a beautiful city. I thank
in particular Leonardo Castellani for his interest and efficient
help. Partial support is also acknowledged from DGAPA-UNAM
project IN119799 and CONACYT project 32307-E.
%\bibliographystyle{plain}
%\bibliography{strings}
%%%%%%%%%%%%%%%%%%%%%%%%%%%%%%%%%%%%%%%%%%%%%%%%%%%%%%%%%%%%%%%%%

%%%%%%%%%%%%%%%%%%%%%%%%%%%%%%%%%%%%%%%%%%%%%%%%%%%%%%%%%%%%%%%%%
%%%%%%%%%%%%%%%%%%%%%%%%%%%%%%%%%%%%%%%%%%%%%%%%%%%%%%%%%%%%%%%%%
\end{document}